\documentclass[12pt,draftclsnofoot,onecolumn]{IEEEtran}
\usepackage{cite}
\usepackage{graphicx}
\usepackage[cmex10]{amsmath}
\usepackage{amssymb}
\usepackage{graphics}
\usepackage{bigints}
\usepackage{amsthm}
\usepackage{color}
\usepackage{epsfig,psfrag}
\usepackage{epstopdf}
\graphicspath{{Figuras/}}
\usepackage{setspace}
\usepackage{amsmath,amssymb,graphicx,float}
\usepackage{dsfont,bbold}

\newcommand\myeq{\mathrel{\overset{\makebox[0pt]{\mbox{\normalfont\tiny\sffamily (a)}}}{=}}}

\begin{document}

\title{{\Large On the Exact Distribution of Mutual Information of Two-user MIMO MAC Based on Quotient Distribution of Wishart Matrices}}

\author{Gabriel Pivaro, Santosh Kumar, and Gustavo Fraidenraich

\thanks{G. Pivaro is at Rice University (USA), as part of his PhD in a joint agreement between Unicamp (Brazil) and Rice University (USA) (email: gabrconc@decom.fee.unicamp.br). Santosh Kumar is with the Department of Physics, Shiv Nadar University, India (email: skumar.physics@gmail.com; santosh.kumar@snu.edu.in). G. Fraidenraich is with the Department of Communications, State University of Campinas (Unicamp), 13083-852, Campinas, SP, Brazil (email:gf@decom.fee.unicamp.br). G. Pivaro and G. Fraidenraich were supported in part by Capes and CNPq, Brazil.}

}
\date{\vspace{-5ex}}

\maketitle

\doublespacing

\begin{abstract}

We propose the exact calculation of the probability density function (PDF) and cumulative distribution function (CDF) of mutual information (MI) for a two-user MIMO MAC network over block Rayleigh fading channels. So far the PDF and CDF have been numerically evaluated since MI depends on the quotient of two Wishart matrices, and no closed-form for this quotient was available. 

We derive exact results for the PDF and CDF of extreme (the smallest/the largest) eigenvalues. Based on the results of quotient ensemble the exact calculation for PDF and CDF of mutual information is presented via Laplace transform approach and by direct integration of joint PDF of quotient ensemble's eigenvalues. Furthermore, our derivations also provide the parameters to apply the Gaussian approximation method, which is comparatively easier to implement. We show that approximation matches the exact results remarkably well for outage probability, i.e. CDF, above 10\%. However, the approximation could also be used for 1\% outage probability with a relatively small error. 

We apply the derived expressions to analyze the effects of adding receiving antennas on the receiver's performance. By supposing no channel knowledge at transmitters and successive decoding at receiver, the capacity of the first user increases and outage probability decreases with extra antennas, as expected.

\end{abstract}

\vspace{0.2cm}
\begin {IEEEkeywords}
Multiple access channel, multiple-input multiple-output, mutual information, outage probability, Rayleigh fading, Wishart matrices, quotient ensemble, extreme eigenvalues, gap probabilities.
\end{IEEEkeywords}

\section{Introduction}


\IEEEPARstart{I}{}t is now well acknowledged that the use of multiple-input multiple-output (MIMO) scheme is crucial to increase the capacity and reliability of wireless systems. MIMO setup provides several benefits such as higher received power via beamforming, higher channel capacity via spatial multiplexing without increasing bandwidth or transmission power, and improved transmission robustness via diversity coding \cite{Gamal2011}. Current cellular systems such as 4G Long Term Evolution (LTE) are using MIMO and the next generation system such as 5G consider the deployment terminal with dozens of antennas, the so-called Massive MIMO.

In this paper, we derive exact expressions to obtain the distribution and the outage probability of the mutual information for a two-user MIMO Multiple Access Channel (MAC). To the best of our knowledge, no exact expressions were derived before, under the following assumptions: channel state information at receiver only (CSIR), Rayleigh fading, and successive decoding \cite{tse2005fundamentals}. The main difficulty to derive exact expressions for this scenario is that the mutual information is a random variable that depends on the quotient of two Wishart matrices \cite{James1960a}.

The recent derivation of the join probability density function (JPDF) of the eigenvalues of the quotient ensemble presented in \cite{Kumar2015} opened the possibility to describe in an exact manner the behavior of the mutual information in this common scenario in wireless networks. Therefore, we derive the exact expressions related to the mutual information that allows, for example, the analysis of the impact of adding more antennas at the receiver (base station) on the performance of network. In addition, we also work out the probability distributions and densities of extreme eigenvalues of the quotient ensemble.

We emphasize that the scenario proposed here is of practical interest since when there is no channel state information at the transmitter (CSIT), then the transmitter encode its messages with a fixed rate \cite{Gamal2011,Goldsmith03capacitylimits}. However, under the slow fading scenario with Rayleigh distribution, the signal transmitted could not be properly decoded at the receiver. In this case, an outage event occurs \cite{Gamal2011}. Our aim here, is to track the outage probability based on the message rate, on the number of antennas at all nodes, and the signal power. 

We assume a successive interference cancellation (SIC) scenario \cite{tse2005fundamentals}, where the first user to be decoded is affected for the signal of the second user (that experience an interference free scenario). Therefore, we focus on the mutual information distribution of the first user. With our expressions, we can quantify the performance improvement achieved with extra power or antennas.

\subsection{Related Works}

The possible application of MIMO in wireless systems probably gained much more attention after Telatar's canonical work \cite{Telatar1999a}. Telatar has shown that the capacity of a MIMO system is directly related to the realizations of the random channel matrix. These realizations are characterized by the probability density function (PDF). However, since the matrix dimensions grow as the number of antennas in the system is increased, evaluation of the capacity is a complex task. One key contribution of Telatar's work was to use random matrix theory (RMT) to show that instead of working with the matrix PDF's, the mutual information distribution could be accessed just by using the JPDF of its eigenvalues. This is possible because of the invariant nature of the mutual information expression under unitary conjugation.

Wang and Giannakis \cite{Wang2004a} showed that the mutual information could be well approximated by a Gaussian distribution. Since a Gaussian distribution is fully characterized by its mean and variance, the problem reduces to working out these two parameters. The calculation of mean of mutual information yields the ergodic capacity, while the Gaussian approximation of mutual information can be used to obtain the outage probability.

The case of a single user MIMO channel has been extensively studied. In \cite{Hyundong2003} a closed-form expression for ergodic capacity was derived for any number of transmit and receive antennas for Rayleigh fading. The exact distribution of mutual information was presented in \cite{Smith2004} for dual MIMO systems under Rician fading. In \cite{Fraidenraich2007a} the MIMO channel capacity over the Hoyt fading channel was presented. In \cite{Kumar2010} random matrix model for the Nakagami-$q$ (Hoyt) fading MIMO communication channels with arbitrary number of transmitting and receiving antennas is considered. The Gaussian approximation was investigated in \cite{Taricco2013} for the Rician fading channel in the asymptotic regime of large number of transmitting and receiving antennas. In \cite{Shang2013} the authors showed that Gaussian approximation remains quite robust even for large signal-to-noise ratio (SNR) for the case of unequal numbers of transmitting and receiving antenna arrays, while it deviates strongly from the exact result for equal number of antenna arrays.

Beside the single user multiple channel scenario mentioned above, MIMO systems have been studied in a variety of multiuser networks such as, Broadcast Channel (BC), Interference Channel (IC), MAC, and Relay Channel. Earlier, much effort was devoted to extend the already known results for single antenna case to the MIMO case \cite{Goldsmith03capacitylimits}. Recent works are investigating how multiple antennas can be utilized to reduce interference in multiuser scenarios \cite{Ghosh2010}.

Although much investigation has been conducted to determine the capacity region and ergodic capacity of MIMO MAC network, only a few works have focused on the determination of outage probability for this channel. In \cite{Ghosh2010} the authors, assuming correlated Rayleigh fading in a multiuser MIMO beam forming network with channel distribution information (CDI), derived a closed-form expression for the outage probability. This expression was used to derive algorithms for joint transmit/receive beam forming and power control to minimize the weighted sum power in the network while guaranteeing this outage probability. In \cite{Poor2012} the authors derived closed-form expressions for outage probability in MIMO IC under the assumption of Gaussian-distributed CSI error, and derived the asymptotic behavior of the outage probability as a function of several system parameters based on the \textit{Chernoff} bound. In \cite{Choi2004} the authors compared the performance in terms of capacity and maximum throughput, of a BC multiuser MIMO system and a MIMO time-division multiple-access (TDMA) MIMO system. Their key assumption is that the number of transmit antennas is much larger than the number of receive antennas at each user and complete knowledge of the channel at the transmitter (CSIT). In \cite{Hunger2009} the authors analyzed asymptotic weighted sum rate maximization in the MIMO multiple access channel. In \cite{Negro2010} the authors propose an iterative algorithm to design optimal linear transmitters and receivers in a $K$-user frequency-flat MIMO IC with CSITR.

\subsection{On the Paper Contributions}

The \textbf{key} contributions of this paper are to obtain exact results for the cumulative distribution function (CDF) and PDF of (i) the extreme eigenvalues of the quotient ensemble comprising two Wishart matrices, and (ii) mutual information for the case when it is a random variable, and again depends on the quotient of two Wishart matrices.

As we show in Sec. II, both the PDF and CDF of mutual information could be written as a function of JPDF, as in the single user case. We invoke the closed-form of JPDF of eigenvalues for the quotient ensemble derived in \cite{Kumar2015} in Sec. III. In Sec. IV we derive closed-form expressions of CDF and PDF for the extreme eigenvalues. With the aid of JPDF, we propose two different methods in Sec. V to derive the exact expressions for PDF and CDF of the mutual information. The first one relies on direct integration of the JPDF, while the second one is to use Laplace transform approach. 

Although the exact expressions for PDF and CDF of the mutual information involve integrals, they provide analytical exact results. Besides the two exact solutions indicated above, we also present in Sec. V the means to obtain the mean and variance of mutual information using the first order and the second order marginal densities (one point and two point correlation functions). With these parameters, we also obtain the Gaussian approximation that is straightforward to use and matches the exact results. We characterize the possible outage values where the approximation matches the exact results extremely well.

Finally, we use the above derivations to analyze the outage probability for a two-user MIMO MAC in a low SNR scenario. The numerical results show that increasing the number of antennas at the base station (BS) decreases the outage probability. The results are evaluated in Sec. VI, where Monte Carlo simulations show perfect agreement with all our analytical expressions.

\section{System Model, Mutual Information Probability Density and Outage Probability}
\label{System Model}

In this section, we first describe the system model under consideration -- the two-user MIMO MAC, a common network that usually appears in the uplink of a cellular-type system \cite{Goldsmith03capacitylimits}. To understand how much information this two-user MIMO MAC could convey, we need to characterize its mutual information. Since, the mutual information is a random variable that depends on realizations of the channel matrix, our goal here is to express its PDF in function of channel matrices' eigenvalues, which reduces the complexity of the problem. Finally, we define the outage probability, that is the mutual information CDF and our main metric to analyze the performance of the two-user MIMO MAC. These expressions are the starting point to derive the exact results proposed in this work.

\subsection{System Model}

\begin{figure}
\centering
\includegraphics[scale=0.35]{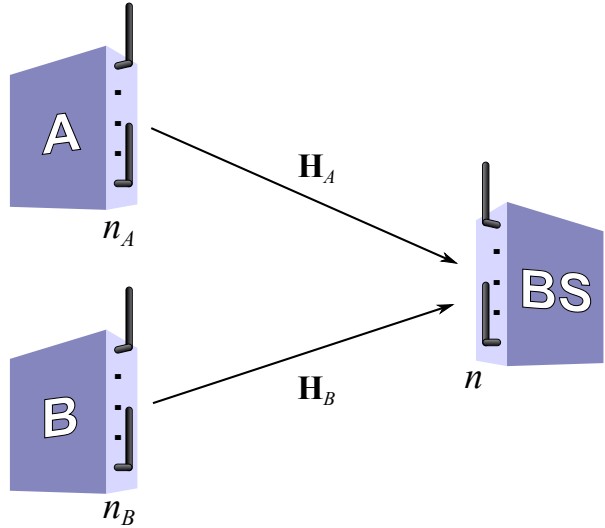}
\caption{System model of two-user MIMO MAC network. User equipment or mobile station A and B have $n_i$, $i=A, B$, transmitting antennas, respectively. The receiver or BS has $n$ receiving antennas. The random channel gain matrix of each user is represented by $\mathbf{H}_i$.}
\label{MAC2User}
\end{figure}

Consider the two-user MIMO MAC network depicted in Fig.~\ref{MAC2User}. The base station (BS) has $n$ receiving antennas and each of the users' equipment or mobile stations has $n_i$, $i=A,\ B$, transmitting antennas. 

The users transmit $\mathbf{u}_i \in \mathbb{C}^{n_i\times1}$, that is circularly symmetric complex Gaussian vector with zero-mean and positive definite covariance matrices $\mathbf{Q}_i$. Users are subject to an individual power constraint of $\text{tr}(\mathbf{Q}_i)\leq n_i$, where $\text{tr}(\cdot)$ is the trace of a matrix. Let $\mathbf{v} \in \mathbb{C}^{n\times1}$ denote the received signal at BS. The $\mathbf{w} \in \mathbb{C}^{n\times1}$ is the noise vector circularly symmetric complex Gaussian with zero-mean and covariance matrix $\mathbf{I}_n$, where $\mathbf{I}_n$ is the $n \times n$ identity matrix. The $n \times n_i$ dimensional channel matrix is denoted by $\mathbf{H}_i$ and its entries are independent and identically distributed (i.i.d.) Gaussian random variables with zero-mean and unit variance. 

The received signal at BS is given by
\begin{equation}
\mathbf{v}=\sqrt{a}\mathbf{H}_A\mathbf{u}_A+\sqrt{b}\mathbf{H}_B\mathbf{u}_B+\mathbf{w}.
\end{equation}
where ``$a=\text{SNR}_A/n_A$ and $b=\text{SNR}_B/n_B$, and $\text{SNR}_i$ are the normalized power ratios of $\mathbf{u}_i$ to the noise (after fading) at each receiver antenna of BS" as stated in \cite{Host-Madsen2005}.

\subsection{Mutual Information}

The BS wishes to recover $\mathbf{u}_i$ from $\mathbf{v}$. Since $\mathbf{u}_i$ and $\mathbf{v}$ are random variables, we use the mutual information to measure how much information BS is able to recover. Then, the MIMO MAC capacity region, assuming successive decoding, is given in terms of mutual information of A and B as \cite{Goldsmith03capacitylimits,tse2005fundamentals} 
\begin{align}
\mathcal{I}_A&= \log_2 \left[\det \left( \mathbf{I}_n + \mathbf{A} + \mathbf{B} \right)  \right] - \mathcal{I}_B\nonumber\\
&= \log_2 \left[\det \left( \mathbf{I}_n + (\mathbf{I}_n + \mathbf{B})^{-1}  \mathbf{A} \right)  \right],
\label{rate:RA}
\end{align}
and
\begin{equation}
\mathcal{I}_B= \log_2 \left[\det \left( \mathbf{I}_n + \mathbf{B}  \right)  \right],
\label{rate:RB}
\end{equation}
where $\mathbf{A}=a\mathbf{H}_A\mathbf{H}_A^\dagger$, $\mathbf{B}=b\mathbf{H}_B\mathbf{H}_B^\dagger$, $\dagger$ denotes the conjugate transpose, and $\det(\cdot)$ is the determinant of a square matrix. Applying similar procedure presented in \cite{Telatar1999a}, we rewrite \eqref{rate:RA} in function of the eigenvalues $\lambda_j$, $j=1,\dots,n$ of the $n \times n$ complex matrix $\mathbf{W}$ as
\begin{equation}
\mathcal{I}_A=\log_2 \det\left(\mathbf{I}_n+\mathbf{W}\right)=\sum_{j=1}^n\log_2(1+\lambda_j),
\label{rate:RA:2}
\end{equation}
where
\begin{equation}
\mathbf{W}=(\mathbf{I}_n + \mathbf{B})^{-1}  \mathbf{A}= (\mathbf{I}_n+b\mathbf{H}_B\mathbf{H}_B^\dagger)^{-1}(a\mathbf{H}_A\mathbf{H}_A^\dagger).
\label{W:quotient}
\end{equation}

Note that we have assumed without loss of generality that BS decodes A's signal first and then B's signal. In this case, the rate of A is affected by the interference caused by B's signal, which does not happen with B \cite{tse2005fundamentals}. In this case, \eqref{rate:RB} is the mutual information of a single-user MIMO channel and is characterized in \cite{Telatar1999a,Wang2004a}. On the other side, the mutual information for the MIMO MAC sum-rate ($\mathcal{I}_A + \mathcal{I}_B$) is given in \cite{Pivaro2014}.

\emph{Therefore, in this work, we focus on the distribution and outage of mutual information of user A given in} \eqref{rate:RA:2}.

\subsection{Outage Probability and Outage Rate}

Now, let us characterize the outage probability and outage rate, the metrics we chose to analyze the performance of the MIMO MAC network. 

We consider in this work, the \emph{slow fading} scenario. In slow fading, with no CSIT, the transmitter encodes $\mathbf{u}_i$ with a fixed rate $R$ bits/s/Hz. An \emph{outage event} could happen when the channel gain is too low for $\mathbf{u}_i$ to be recovered \cite{Gamal2011}. The probability of occurrence of an outage event is known as outage probability, and is given by \cite[Eq. (5.54)]{tse2005fundamentals}:
\begin{align}
p_{out}(R)&=\text{Pr} \left\lbrace \mathcal{I}_A  < R \right\rbrace\nonumber\\
&=\text{Pr} \left\lbrace \log_2 \left[\det \left( \mathbf{I}_n + \mathbf{W}\right) \right]  < R \right\rbrace.
\label{p:out:1}
\end{align}

The outage rate is defined in \cite{Wang2004a} as the rate $R$ for which the outage probability is at the given level $\varepsilon$:
\begin{equation}
R_{out} = \arg_R [p_{out}(R) = \varepsilon].
\end{equation}
In other words, the outage rate is the rate conveyed subject to outage probability equal to $\varepsilon$ \cite{Gamal2011}.

Since working with $\mathbf{W}$ is not straightforward because the number of integrals is related with the number of $\mathbf{W}$'s entries, we adopt similar procedure from \cite{Telatar1999a} and rewrite \eqref{p:out:1} as a function of the eigenvalues of $\mathbf{W}$. Then, the outage probability is given by\begin{align}
\label{Pout}
p_{out}(R)&=
\text{Pr} \left\lbrace \left( \prod_{i=1}^n ( 1 + \lambda_j) \right)   < 2^R \right\rbrace\nonumber\\
&\myeq\int_0^\infty \cdots \int_0^\infty \Theta\left(2^R- \prod_{j=1}^n ( 1 + \lambda_j) \right)P(\lambda_1,...,\lambda_n)\ d\lambda_{1}\dots d\lambda_{n}\nonumber\\
&=\int_0^\infty \cdots \int_0^\infty \Theta\left(R - \sum_{j=1}^n \log_2( 1 + \lambda_j) \right)P(\lambda_1,...,\lambda_n)\ d\lambda_{1}\dots d\lambda_{n},
\end{align}
where $\Theta(\cdot)$ represents the Heaviside-theta function, with $\Theta(x)=0$ for $x<0$ and $\Theta(x)=1$ for $x>0$, $x \in \mathbb{R}$. Note that  (a) follows because the theta function ensures that contribution to the probability comes only from the region where $\prod_{i=1}^n ( 1 + \lambda_j) < 2^R$. Evidently, determining the outage probability amounts to calculating the CDF of the mutual information.

Finally, the PDF of mutual information is obtained by differentiating \eqref{Pout} and is given by
\begin{equation}
\label{PMI}
p(\mathcal{I}_A)=\frac{dp_{out}(x)}{dx}\Big|_{x=\mathcal{I}_A}=\int_0^\infty \cdots\int_0^\infty  \delta\,\Big(\mathcal{I}_A-\sum_{j=1}^n \log_2(1+ \lambda_j)\Big) P(\lambda_1,...,\lambda_n)\ d\lambda_1 \cdots d\lambda_n,
\end{equation}
where $\delta(\cdot)$ is the Dirac-delta function \cite[pg. 1029]{abramowitz2012handbook}.
\vspace{0.5cm}

The expressions \eqref{Pout} and \eqref{PMI} are the formal solutions to the outage probability and the density of mutual information. In Section IV we plug the JPDF in \eqref{Pout} and \eqref{PMI} and present the final expressions. We also present an alternative form based on Laplace Transform which is also exact, and represent an alternative in terms of computation time. Finally, a Gaussian approximation is also presented. This last solution provides a trade-off between accuracy and time.

\section{The Quotient Ensemble Eigenvalues Distribution}

In the previous section, we showed that the PDF and CDF of mutual information depends on the JPDF $P(\lambda_1,...,\lambda_n)$ of $\mathbf{W}$. In this section, we invoke the recently derived JPDF $P(\lambda_1,...,\lambda_n)$ for a quotient comprising Wishart matrices. We  link this result to the $r$-point correlation function. Both the JPDF $P(\lambda_1,...,\lambda_n)$ of $\mathbf{W}$ and the $r$-point correlation function will be used in the following sections to derive our proposed expressions.
\vspace{0.5cm}

Consider the quotient ensemble of random matrices $\mathbf{W}$ as defined in \eqref{W:quotient}. The probability density of these $n\times n$ dimensional complex matrices was recently derived in~\cite{Kumar2015}:
\begin{equation}
p(\mathbf{W})\propto e^{-a^{-1}\text{tr}\mathbf{W}} \det(\mathbf{W})^{n_A-n}\,\Psi\big(n_B,n_A+n_B+n;(b^{-1}\mathbf{I}_n +a^{-1}\mathbf{W})\big).
\end{equation}
Here $\Psi(\cdot)$  is the confluent hypergeometric function of the second kind (\textit{Tricomi} function) with matrix argument~\cite{JoshiJoshi1985}:
\begin{equation} 
\Psi(\alpha,\gamma;\mathbf{X})=\frac{1}{\pi^{n(n-1)/2}\prod_{j=1}^n\Gamma(\alpha-j+1)}\int_{ \mathbf{Y}>0} d\mathbf{Y} e^{-\text{tr} (\mathbf{XY})}| \mathbf{Y}|^{\alpha-n}|\mathbf{I}+ \mathbf{Y}|^{\gamma-\alpha-n},
\end{equation}
with Re$(\mathbf{X})>0$, Re$(\alpha)>(n-1)$ for convergence, and Re$(\cdot)$ denotes the real part.

The JPDF $P(\lambda_1,...,\lambda_n)$ of eigenvalues of $\mathbf{W}$ exhibits a biorthogonal structure of Borodin type~\cite{Borodin1998704}, and is given by~\cite{Kumar2015}
\begin{equation}
\label{biortho}
P(\lambda_1,...,\lambda_n)=C_n \Delta_n(\{\lambda\})\prod_{i=1}^n e^{-\lambda_i/a}\,\lambda_i^{n_A-n} \det\left[f_j(\lambda_k)\right]_{j,k=1,...,n},
\end{equation}
where
\begin{equation*}
\Delta_n(\{\lambda\})=\det[\lambda_k^{j-1}]_{j,k=1,...,n} =\prod_{j>k}(\lambda_j-\lambda_k),
\end{equation*}
is the Vandermonde determinant, and
\begin{equation*}
f_j(\lambda_k)=U\Big(n_B -j+1,n_A+n_B-j+2;~\frac{1}{b}+\frac{\lambda_k}{a}\Big)
\end{equation*}
is in terms of the usual confluent hypergeometric function $U(\cdot)$ of the second type (\textit{Tricomi} function).\footnote{To avoid any confusion we have used distinct symbols to represent confluent hypergeometric function with matrix argument ($\Psi$), and that with scalar argument ($U$).}

The normalization factor in~\eqref{biortho},  $C_n$, turns out to be
\begin{align}
\nonumber
C_n^{-1}&=n!\,\det[h_{j,k}]_{j,k=1,...,n} \\
 &=n!\, a^{n n_A-n (n-1)/2}\,b^{n n_B}\prod_{j=1}^n \Gamma(j)\Gamma(n_A-j+1),
\end{align}
with
\begin{align}
\label{hjk}
\nonumber
h_{j,k}&=\int_0^\infty  e^{-\lambda/a}\,\lambda^{n_A-n+k-1} f_j(\lambda)\ d\lambda\\
&=a^{n_A-n+k}\Gamma(n_A-n+k)\, U\Big(n_B-j+1,n_B+n-j-k+2;~\frac{1}{b}\Big).
\end{align}

The $r$-point correlation function \cite{Mehta2004}, $(1\leq r\leq n)$, corresponding to \eqref{biortho} is given by~\cite{Kumar2015}:
\begin{align}
\label{corrfunc}
R_r(\lambda_1,...,\lambda_r)&=\frac{n!}{(n-r)!}\int_0^\infty \cdots \int_0^\infty P(\lambda_1,...,\lambda_n)\   d\lambda_{n} \cdots d\lambda_{r+1}\nonumber\\
&=(-1)^r n! \,C_n \prod_{l=1}^r  e^{-\lambda_l/a}\,\lambda_l^{n_A-n} \, \det\begin{bmatrix}   \mathbb{0} &  [\lambda_j^{k-1}]_{\substack{j=1,...,r\\k=1,...,n}}  \\  [f_j(\lambda_k)]_{\substack{j=1,...,n\\k=1,...,r}}  & [h_{j,k}]_{\substack{j=1,...,n\\k=1,...,n}} \end{bmatrix}
\end{align}
where $\mathbb{0}$ represents a $r\times r$ block with all entries zero. 

The one-point function $R_1(\lambda_1)$  and the two-point function $R_2(\lambda_1,\lambda_2)$ will be useful in order to obtain the Gaussian approximation. We note that the one-point function is related to the marginal density as $p_1(\lambda)=R_1(\lambda)/n$, while the two-point function gives the JPDF of two eigenvalues as $p_2(\lambda_1,\lambda_2)=R_2(\lambda_1,\lambda_2)/(n(n-1))$.

\section{Extreme eigenvalues statistics}

Along with the mutual information PDF and CDF that depends of the distribution of all $n$ eigenvalues\footnote{With exception for the Gaussian approximation case that will be show in Section \ref{Proposed Mutual Information Exact Distribution and Outage Probability}.} as shown in \eqref{Pout} and \eqref{PMI}, respectively, the distribution of the extreme eigenvalues (the smallest/the largest) \cite{Khatri1964} also serve as important metric for analyzing the performance of MIMO systems \cite{Kang2003,Zanella2008,Zanella2008b}. 

In this section, we derive exact results for the distributions and densities of both the smallest eigenvalue ($\lambda_\mathrm{min}$) and the largest eigenvalue ($\lambda_\mathrm{max}$) of the quotient ensemble defined in \eqref{W:quotient}. These are based on the general results summarized in~\cite{Kumar2015a}. We first present exact results for the gap probability which refers to the probability of finding no eigenvalue in a given interval. These are then used to obtain the densities of $\lambda_\mathrm{min}$ and $\lambda_\mathrm{max}$.

The probability that there are no eigenvalues between 0 and $x$, or equivalently the probability that all eigenvalues are greater than or equal to $x$ is given by
\begin{equation}
E((0,x))= \int_x^\infty \cdots\int_x^\infty  P(\lambda_1,\dots,\lambda_n)\ d\lambda_1\dots d\lambda_n.
\end{equation}
On the other hand we have
\begin{equation}
E((x,\infty))=\int_0^x \cdots\int_0^x   P(\lambda_1,\dots,\lambda_n)\ d\lambda_1 \dots d\lambda_n,
\end{equation}
which gives the probability that there are no eigenvalue between $x$ and $\infty$, or equivalently that all eigenvalues are less than or equal to $x$. Inserting the JPDF given in \eqref{biortho} in the above equation and implementing \textit{Andr\'{e}ief's} integration formula~\cite{Andreief1883}, at once yield the result for the above gap probabilities in the present case. We have
\begin{equation}
\label{E1}
E((0,x))=n! \,C_n \det[\chi_{j,k}((0,x))]_{j,k=1,...,n},
\end{equation}
with the kernel $\chi_{j,k}((0,x))$ given by
\begin{align}
\chi_{j,k}((0,x))&=\int_x^\infty   e^{-\lambda/a}\,\lambda^{n_A-n+k-1} f_j(\lambda)\ d\lambda \nonumber\\
&=e^{-x/a}\sum_{r=0}^{n_A-n+k-1} \frac{\Gamma(n_A-n+k)}{\Gamma(n_A-n+k-r)}a^{r+1}x^{n_A-n+k-r-1}\,\nonumber\\ &\times U\Big(n_B-j+1,n_A+n_B-j-r+1;\frac{1}{b}+\frac{x}{a}\Big).
\end{align}

To obtain the finite-sum result in the second line above, we used the transformation $\mu= \lambda-x$, applied the binomial expansion on the resulting factor $(\mu+x)^{n_A-n+k-1}$, and finally performed term by term integration over $\mu$. We note that for $x\rightarrow 0$, $\chi_{j,k}((0,x))$ reduces to $h_{j,k}$ as in~\eqref{hjk}.
In a similar way we obtain
\begin{equation}
\label{E2}
E((x,\infty))=n! \,C_n \det[\chi_{j,k}((x,\infty)]_{j,k=1,...,n},
\end{equation}
where $\chi_{j,k}((x,\infty)$ is given by
\begin{align}
&\chi_{j,k}((x,\infty))=\int_0^x   e^{-\lambda/a}\,\lambda^{n_A-n+k-1} f_j(\lambda)\ d\lambda =h_{j,k}-\chi_{j,k}((0,x))
\end{align}

Now, since $E((0,x))$ gives the survival function\footnote{Survival function and cumulative distribution function are related as SF$=1-$CDF.} (SF) or reliability function of the $\lambda_\mathrm{min}$, it can be used to obtain the corresponding PDF. It is given by
 \begin{align}
\label{pS}
p_{\lambda_\mathrm{min}}(x)&=-\frac{d}{dx}E((0,x))=n!\,C_n\,\sum_{i=1}^n \det[\phi_{j,k}^{(i)}(x)]_{j,k=1,...,n},
\end{align}
where
\begin{equation}
\phi_{j,k}^{(i)}(x)
=\begin{cases}
 e^{-x/a}\,x^{n_A-n+k-1} f_j(x), & j=i,\\
 \chi_{j,k}((0,x)), & j\neq i.
 \end{cases}
 \end{equation}
Similarly, $E((x,\infty))$ is the cumulative distribution function (CDF) of the $\lambda_\mathrm{max}$, and hence the PDF of $\lambda_\mathrm{max}$ is obtained as
\begin{align}
\label{pL}
p_{\lambda_\mathrm{max}}(x)&=\frac{d}{dx}E((x,\infty))=n!\,C_n\,\sum_{i=1}^n \det[\psi_{j,k}^{(i)}(x)]_{j,k=1,...,n},
\end{align}
with
\begin{equation}
\psi_{j,k}^{(i)}(x)
=\begin{cases}
 e^{-x/a}\,x^{n_A-n+k-1} f_j(x), & j=i,\\
 \chi_{j,k}((x,\infty)), & j\neq i.
 \end{cases}
 \end{equation}

We show in Section \ref{NumRes} that the above exact expressions agree perfectly with the Monte Carlo simulations.

\section{Proposed Mutual Information Exact Density and Outage Probability}
\label{Proposed Mutual Information Exact Distribution and Outage Probability}

In this section, we present two exact ways to obtain the PDF and outage probability of mutual information. Moreover, the Gaussian approximation is also presented, since it leads to reasonable results and is more straightforward than the exact solutions. 

\subsection{Exact results based directly on JPDF}

We first calculate the PDF of mutual information. For this we notice that one of the integrals in \eqref{PMI} can be easily performed because of the Dirac-delta function and leaves us with 

\begin{align}
\label{PMI2}
p(\mathcal{I}_A)&=\ln2\int_0^\infty \cdots\int_0^\infty \frac{2^{\mathcal{I}_A}}{\prod_{j=2}^n (1+\lambda_j)}\ P\left( \frac{2^{\mathcal{I}_A}}{\prod_{j=2}^n(1+\lambda_j)}-1,\lambda_2,...,\lambda_n\right) \nonumber\\
&~~~~~\times\Theta\left(2^{\mathcal{I}_A}-\prod_{j=2}^n(1+\lambda_j)\right)\ d\lambda_n \cdots d\lambda_2\nonumber\\
&=\ln2\int_0^{u_2} \cdots\int_0^{u_n}  \frac{2^{\mathcal{I}_A}}{\prod_{j=2}^n (1+\lambda_j)}\ P\left( \frac{2^{\mathcal{I}_A}}{\prod_{j=2}^n(1+\lambda_j)}-1,\lambda_2,...,\lambda_n\right)\ d\lambda_n \cdots d\lambda_2,
\end{align}
where
\begin{equation}
\nonumber
u_\mu=\frac{2^\mathcal{I}_A(1+\lambda_2)}{\prod_{j=2}^{\mu} (1+\lambda_j)}-1=\frac{2^\mathcal{I}_A}{\prod_{j=2}^{\mu-1} (1+\lambda_j)}-1.
\end{equation}

\emph{Special case} $(n=2)$: we have \cite[Eq. 6.40]{papoulis2002probability}
\begin{equation}
p(\mathcal{I}_A)=\ln2\int_0^{2^{\mathcal{I}_A}-1} \frac{2^{\mathcal{I}_A}}{1+\lambda_2}\ P\left(\frac{2^\mathcal{I}_A}{1+\lambda_2}-1,\lambda_2\right)\ d\lambda_2.
\end{equation}

The outage probability can be written using~\eqref{Pout} as
\begin{align}
p_{out}(R)=\int_0^{v_1} \cdots\int_0^{v_n} P\left( \lambda_1,\lambda_2,...,\lambda_n\right)\ d\lambda_n \cdots d\lambda_1,
\label{p:out:exact:jpdf}
\end{align}
where
\begin{equation}
\nonumber
v_\mu=\frac{2^R(1+\lambda_1)}{\prod_{j=1}^{\mu} (1+\lambda_j)}-1=\frac{2^R}{\prod_{j=1}^{\mu-1} (1+\lambda_j)}-1.
\end{equation}
Note that \eqref{p:out:exact:jpdf} simplifies \eqref{Pout} by transferring the summation of eigenvalues to the limits of the integral. Another possible expression for the outage probability use directly \eqref{PMI2} and can be written as
\begin{multline}
p_{out}(R)=\int_0^R  p(\mathcal{I}_A)\ d\mathcal{I}_A\\
=\ln2\int_0^R \int_0^{u_2} \cdots\int_0^{u_n}  \frac{2^{\mathcal{I}_A}}{\prod_{j=2}^n (1+\lambda_j)}\ P\left( \frac{2^{\mathcal{I}_A}}{\prod_{j=2}^n(1+\lambda_j)}-1,\lambda_2,...,\lambda_n\right)\  d\lambda_n \cdots d\lambda_2\ d\mathcal{I}_A.
\end{multline}

\subsection{Exact results based on Laplace transform approach}

The Laplace transform of $p(\mathcal{I}_A)$ defined in \eqref{PMI} is given by
\begin{equation}
\widetilde{p}(s)=\mathcal{L}[p(\mathcal{I}_A)](s)=\int_0^\infty \cdots\int_0^\infty  \,e^{-s\sum_{j=1}^n \log_2(1+ \lambda_j )} P(\lambda_1,...,\lambda_n)\ d\lambda_1 \cdots d\lambda_n.
\end{equation}
The above expression serves as the moment generating function (MGF) for $\mathcal{I}_A$, since the moments of $\mathcal{I}_A$ can be obtained using the coefficients of powers of $s$ in the series expansion of $\widetilde{p}(s)$. Using the JPDF given in~\eqref{biortho} we obtain
\begin{align}
\nonumber
\widetilde{p}(s)=C_n\int_0^\infty \cdots\int_0^\infty   \Delta(\{\lambda\})\prod_{l=1}^n e^{-\lambda_l/a}\lambda_l^{n_A-n}(1+\lambda_l)^{-(s/\ln 2)}\ d\lambda_1 \cdots d\lambda_n \\
\times\det\Big[U\Big(n_B -j+1,n_A+n_B-j+2;~\frac{1}{b}+\frac{\lambda_k}{a}\Big)\Big]_{j,k=1,...,n},
\end{align}
where we used $\log_2 z=\ln z/\ln2$. 

With the aid of \textit{Andr\'{e}ief's} integration formula~\cite{Andreief1883} the above result can be immediately cast in the form of a determinant
\begin{equation}
\label{LTMI}
\widetilde{p}(s)=n!\, C_n\,\det[\psi_{j,k}(s)]_{j,k=1,...,n},
\end{equation}
where 
\begin{equation}
\psi_{j,k}(s)=\int_0^\infty (1+\lambda)^{-(s/\ln2)}\lambda^{n_A-n+k-1} e^{-\lambda/a}\, U\Big(n_B -j+1,n_A+n_B-j+2;~\frac{1}{b}+\frac{\lambda}{a}\Big)\ d\lambda.
\nonumber
\end{equation}

The density of $\mathcal{I}_A$ follows by taking the inverse Laplace of $\widetilde{p}(s)$,
\begin{equation}
\label{Eq:pIA:InverseLaplace}
p(\mathcal{I}_A)=\mathcal{L}^{-1}\{\widetilde{p}(s)\}(\mathcal{I}_A).
\end{equation}

The outage probability follows by taking the inverse Laplace of  $s^{-1}\widetilde{p}(s)$~\cite{Wang2004a},
\begin{equation}
p_{out}(R)=\mathcal{L}^{-1}\{s^{-1}\widetilde{p}(s)\}(R).
\end{equation}

With the above results we may recover the densities by performing numerical inversion of Laplace transform as in \cite{Wang2004a}. However, we make further analytical progress below to 	obtain an alternate expression for the PDF of mutual information.

\emph{Special case} $(n=1)$:
Let us consider the $n=1$ case. The density of mutual information can be obtained from \eqref{PMI} as
\begin{equation}
p(\mathcal{I}_A)=C_1 \int_0^\infty  \delta(\mathcal{I}_A-\log_2(1+\lambda)) e^{-\lambda/a}\lambda^{n_A-1}\,U\left(n_B,n_A+n_B+1;\frac{1}{b}+\frac{\lambda}{a}\right)\ d\lambda.
\nonumber
\end{equation}

This one-dimensional integral can be readily performed because of the presence of Dirac-delta function and yields the exact result for $n=1$ as
\begin{equation}
p(\mathcal{I}_A)=\frac{a^{-n_A}b^{-n_B}}{\Gamma(n_A)} (\ln2)\, 2^\mathcal{I}_A\, (2^\mathcal{I}_A-1)^{n_A-1}\exp\Big(-\frac{2^\mathcal{I}_A-1}{a}\Big) \,U\left(n_B,n_A+n_B+1;\frac{1}{b}+\frac{2^\mathcal{I}_A-1}{a}\right).
\label{p:Ia:int}
\end{equation}

Comparing \eqref{p:Ia:int} with \eqref{LTMI} evaluated for $n=1$, we arrive at the following inverse Laplace transform identity:
\begin{multline}
\mathcal{L}^{-1}\Big[\int_0^\infty \lambda^\gamma e^{-\lambda/a}(1+\lambda)^{-s/\ln2}\,U\Big(\alpha,\beta,\frac{1}{b}+\frac{\lambda}{a}\Big)d \lambda\Big](t)=\nonumber\\
(\ln2)\,2^t\,\exp\left(-\frac{(2^t-1)}{a}\right)(2^t-1)^\gamma\,U\Big(\alpha,\beta,\frac{1}{b}+\frac{2^t-1}{a}\Big).
\end{multline}

With this interesting result in our hands we can use the convolution property of the Laplace transform and write an expression for the PDF of mutual information for arbitrary $n$ as a $(n-1)$ fold integral. To this end we expand the determinant in \eqref{LTMI} and afterwards use the following result for inverse Laplace transform of product of $n$ functions, which follows from the result for product of two functions \cite{Bracewell2000}
\begin{multline}
\mathcal{L}^{-1}[\widetilde{F}_1(s)\widetilde{F}_2(s)\cdots\widetilde{F}_n(s)](x_1)=\\
\int_0^{x_1} \int_0^{x_2} \cdots \int_0^{x_{n-1}} F_1(x_1-x_2) F_2(x_2-x_3)\cdots F_{n-1}(x_{n-1}-x_n)F_n(x_n)\  dx_n\  \cdots dx_3\ dx_2,
\label{Laplace:conv}
\end{multline}
where 
\begin{equation}
\nonumber
\mathcal{L}^{-1}[\widetilde{F}_j(s)](t)=F_j(t),~~~~ j=1,...,n.
\end{equation}

Therefore, with the help of \eqref{Laplace:conv} in  \eqref{Eq:pIA:InverseLaplace}, we obtain the following expression: \begin{multline}
\label{Eq:pIA:general}
p(\mathcal{I}_A)=n!\, C_n\,(\ln 2)^n\,2^{x_1}\, \int_0^{x_1}\int_0^{x_2} \cdots \int_0^{x_{n-1}}\prod_{j>k}(2^{x_j-x_{j+1}}-2^{x_k-x_{k+1}})\\
\times \prod_{j=1}^n(2^{x_j-x_{j+1}}-1)^{n_A-n}\exp\left(-\frac{(2^{x_j-x_{j+1}}-1)}{a} \right)\\
\times U\left(n_B-j+1,n_A+n_B-j+2;\frac{1}{b}+\frac{2^{x_j-x_{j+1}}-1}{a}\right)\ dx_n\cdots dx_3\ dx_2,
\end{multline}
where
$x_1\equiv \mathcal{I}_A$ and $x_{n+1}\equiv 0$. 

\emph{Special case} $(n=2)$: we have the following explicit result
\begin{multline}
p(\mathcal{I}_A)=\frac{a^{1-2n_A }b^{-2n_B}}{\Gamma(n_A)\Gamma(n_A-1)}\,(\ln 2)^2\,2^{\mathcal{I}_A}\, \int_0^{\mathcal{I}_A}(2^{x}-2^{\mathcal{I}_A-x})(2^{\mathcal{I}_A-x}-1)^{n_A-2}(2^{x}-1)^{n_A-2}\\
\times \exp\left(-\frac{(2^{\mathcal{I}_A-x}+2^{x}-2)}{a} \right) U\left(n_B,n_A+n_B+1;\frac{1}{b}+\frac{2^{\mathcal{I}_A-x}-1}{a}\right)\\
\times U\left(n_B-1,n_A+n_B;\frac{1}{b}+\frac{2^{x}-1}{a}\right)\ dx.
\end{multline}

The outage probability, which is the CDF of the mutual information, can be written as the integral of \eqref{Eq:pIA:general} from $0$ to $R$ as
\begin{multline}
p_{out}(R)=\int_0^R p(x_1)\ dx_1\\
=n!\, C_n\,(\ln 2)^n\, \int_0^{R} \int_0^{x_1}\int_0^{x_2} \cdots \int_0^{x_{n-1}} 2^{x_1}\prod_{j>k}(2^{x_j-x_{j+1}}-2^{x_k-x_{k+1}})\\
\times \prod_{j=1}^n(2^{x_j-x_{j+1}}-1)^{n_A-n}\exp\left(-\frac{(2^{x_j-x_{j+1}}-1)}{a} \right)\\
\times U\left(n_B-j+1,n_A+n_B-j+2;\frac{1}{b}+\frac{2^{x_j-x_{j+1}}-1}{a}\right)\ dx_n\cdots dx_3\ dx_2\ dx_1.
\end{multline}

We should remark at this point that for the evaluation of PDF and outage probability of $\mathcal{I}_A$, as far as number of integrals is concerned, we have not gained anything. However, the above exact expressions provide an alternative route to calculate these quantities compared to the expressions derived in the last subsection, where we adopted the strategy of integrating the JPDF directly.

\subsection{Gaussian Approximation}

The expressions for PDF and CDF presented above use the JPDF of the eigenvalues or Laplace transform, which involves the calculation of multiple integrals. A more straightforward method is to use the Gaussian approximation that depends only on integrals involving up to two eigenvalue density.

The PDF of mutual information can be approximated by a Gaussian distribution as \cite{Wang2004a}
\begin{equation}
p(\mathcal{I}_A)\approx \frac{1}{\sqrt{2\pi\sigma_{\mathcal{I}_A}^2}}\exp\left(-\frac{(\mathcal{I}_A-\mu_{\mathcal{I}_A})^2}{2\sigma_{\mathcal{I}_A}^2}\right).
\end{equation}
Correspondingly, the outage probability is given by
\begin{equation}
p_{out}(R)\approx \frac{1}{2}\text{erfc}\left(\frac{\mu_{\mathcal{I}_A}-R}{\sqrt{2\sigma_{\mathcal{I}_A}^2}}\right),
\end{equation}
where erfc$(\cdot)$ represents the complementary error function.

In principle, we can obtain the mean ($\mu_{\mathcal{I}_A}$) and variance ($\sigma_{\mathcal{I}_A}^2$) with the aid of the coefficients of $s$ and $s^2$ in the series expansion of $\widetilde{p}(s)$. However, this is nontrivial because of the complicated structure of $\widetilde{p}(s)$ in \eqref{LTMI}. Therefore, we resort to the strategy of obtaining $\mu_{\mathcal{I}_A}$ and $\sigma_{\mathcal{I}_A}^2$ with the help of the one-point and two-point correlation functions given in \eqref{corrfunc}.

The $\mu_{\mathcal{I}_A}$ can be obtained by averaging over the ensemble of $\mathbf{W}$ as \cite{Telatar1999a}
\begin{align*}
\mu_{\mathcal{I}_A}&=\mathds{E}\big[\mathcal{I}_A(\mathbf{W})\big]\\
&=\mathds{E}\left[\sum_{j=0}^n\log_2(1+\lambda_j)\right]\\
&=n\,\mathds{E}\big[\log_2(1+\lambda_1)\big],
\end{align*}
and using the one-point function $R_1(\lambda_1)$ this can be written as
\begin{equation}
\mu_{\mathcal{I}_A}=\int_0^\infty R_1(\lambda_1)\log_2(1+\lambda_1)\ d\lambda_1.
\end{equation}

Similarly, the $\sigma_{\mathcal{I}_A}^2$ can be obtained as \cite{Wang2004a}
\begin{align}
\nonumber
\sigma^2_{\mathcal{I}_A}&=\mathds{E}\big[\mathcal{I}_A^2(\mathbf{W})\big]-(\mathds{E}\big[\mathcal{I}_A(\mathbf{W})\big])^2\\
\nonumber
&=\mathds{E}\left[\left(\sum_{j=1}^n\log_2(1+\lambda_j)\right)^2\right]-\mu_{\mathcal{I}_A}^2\\
\nonumber
&=\mathds{E}\left[\sum_{j,k=1}^n\log_2(1+\lambda_j)\log_2(1+\lambda_k)\right]-\mu_{\mathcal{I}_A}^2\\
\nonumber
&=\mathds{E}\left[\sum_{j=1}^n\log_2^2(1+\lambda_j)+\sum_{j\neq k}\log_2(1+\lambda_j)\log_2(1+\lambda_k)\right]-\mu_{\mathcal{I}_A}^2\\
&=n\,\mathds{E}\left[\log_2^2(1+\lambda_1)\right]+n(n-1)\,\mathds{E}\left[\log_2(1+\lambda_1)\log_2(1+\lambda_2)\right]-\mu_{\mathcal{I}_A}^2.
\end{align}

Therefore, with the aid of one-point and two-point correlation functions \eqref{corrfunc} this can be written as
\begin{multline}
\sigma_{\mathcal{I}_A}^2=\int_0^\infty  R_1(\lambda_1)\log_2^2(1+\lambda_1)\ d\lambda_1\\
+\int_0^\infty \int_0^\infty  R_2(\lambda_1,\lambda_2)\log_2(1+\lambda_1)\log_2(1+\lambda_2)-\mu_{\mathcal{I}_A}^2\ d\lambda_1\ d\lambda_2.
\end{multline}
Mathematically, we can see that the advantage of using Gaussian approximation is that we need to perform only up to a two-fold integral, instead of $n-1$ or $n$-fold integrals required for exact results.

\begin{figure}
\centering
\includegraphics[scale=0.65]{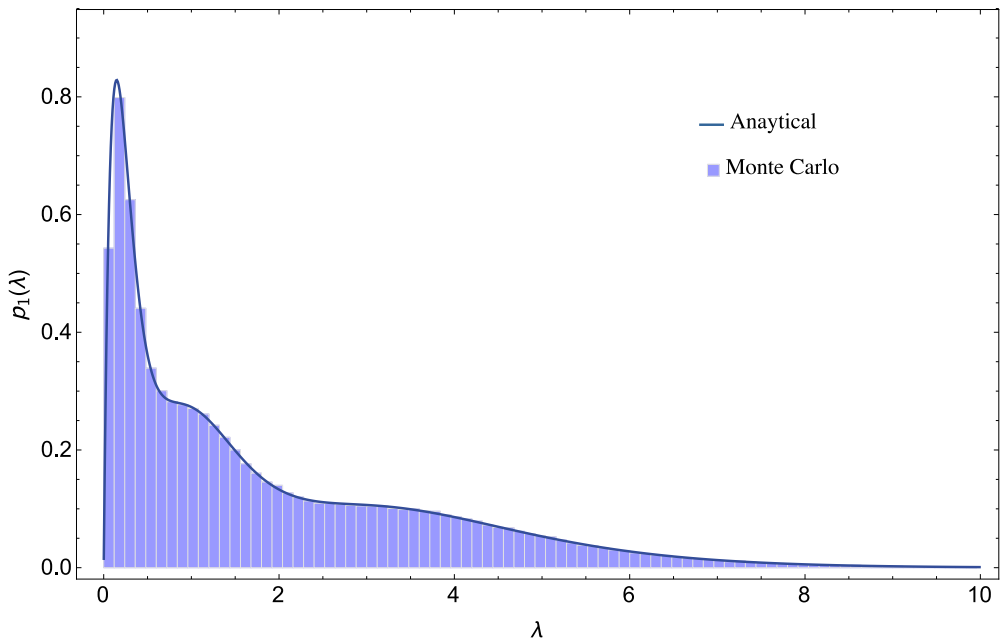}
\caption{Marginal density of eigenvalues of the quotient ensemble. The parameters used are $n=3$, $n_A=4$, $n_B=5$, $a=1$ and $b=1/3$.}
\label{Marginal_Density}
\end{figure}

\begin{figure}[ht!]
\centering
\includegraphics[scale=0.65]{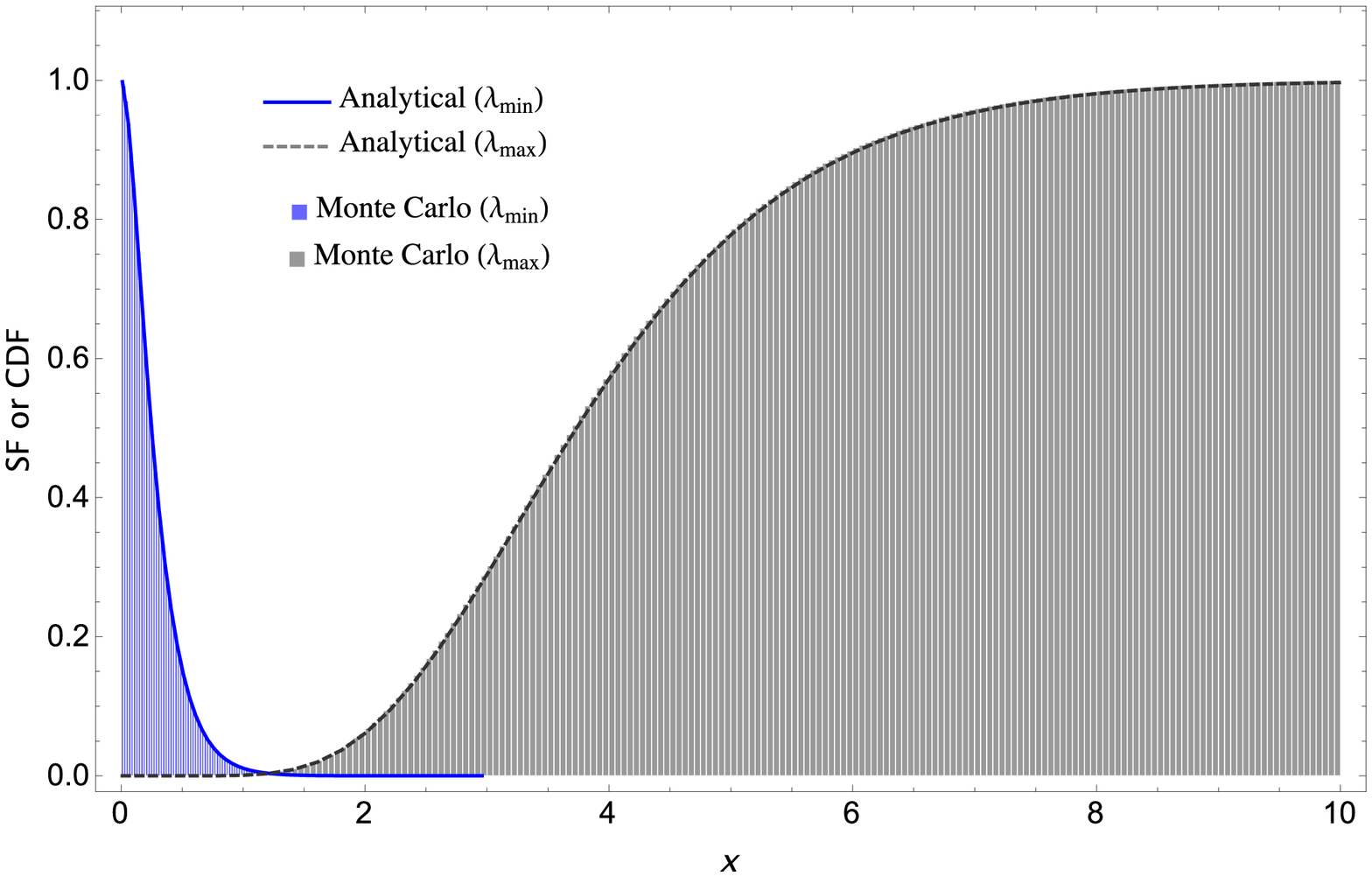}
\caption{SF of the smallest eigenvalue ($\lambda_\mathrm{min}$) and CDF of the largest eigenvalue ($\lambda_\mathrm{max}$), as given by~\eqref{E1} and~\eqref{E2}. The parameter values are $n=3$, $n_A=4$, $n_B=5$, and $b=1/3$.}
\label{Extreme_CDF}
\end{figure}

\begin{figure}[h!]
\centering
\includegraphics[scale=0.65]{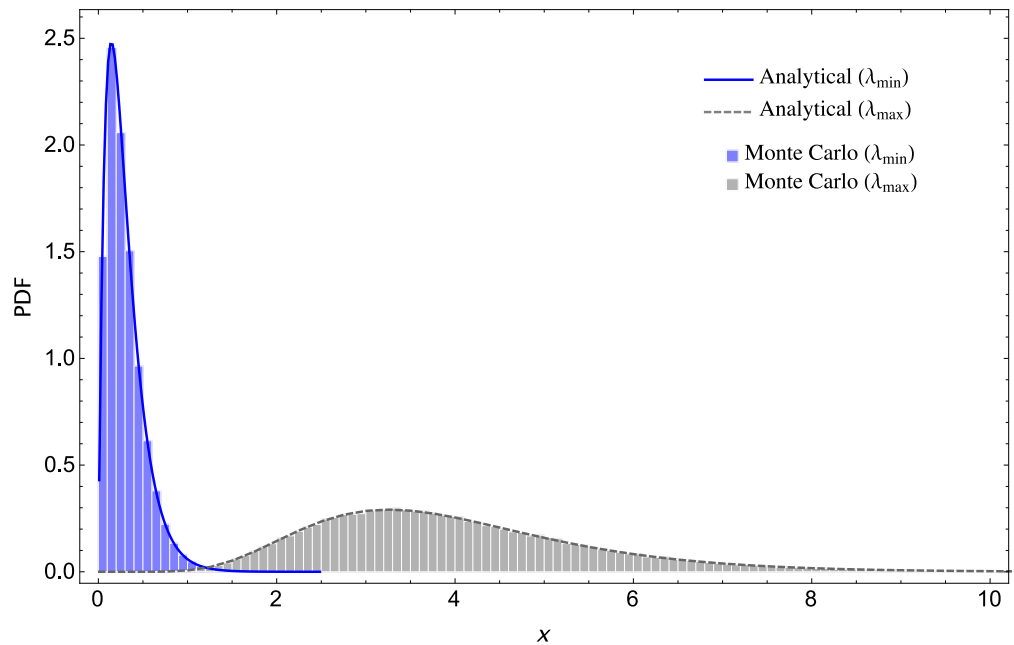}
\caption{PDF of the smallest eigenvalue ($\lambda_\mathrm{min}$) and that of the largest eigenvalue ($\lambda_\mathrm{max}$), as given by~\eqref{pS} and~\eqref{pL}. The parameters values are $n=3$, $n_A=4$, $n_B=5$, and $b=1/3$.}
\label{Extreme_PDF}
\end{figure}

\section{Numerical Results}
\label{NumRes}

In this section we present a numerical example in order to validate the exact expressions proposed in this work. We begin with the comparison of results concerning the eigenvalues of the quotient ensemble, viz., the marginal density, and probability distributions and densities of the extreme eigenvalues. Afterwards, we move over to examine the behavior of the mutual information. As will be shown, there is a perfect match between the results from Monte Carlo simulations and the exact results presented in the preceding sections.

We consider the following scenario. Suppose that user A transmits with $a=1$ ($\text{SNR}_A=n_A\times a=4 \times 1=6.02$ dB), and user B transmits with $b=1/3$ ($\text{SNR}_B=n_B\times b=5 \times 1/3=2.21$ dB). In Fig~\ref{Marginal_Density} we show the marginal density of eigenvalues for $n=3$, while in Figs.~\ref{Extreme_CDF} and~\ref{Extreme_PDF} we display the probability distributions and densities of the extreme eigenvalues. 

\begin{figure}[ht!]
\centering
\includegraphics[scale=0.75]{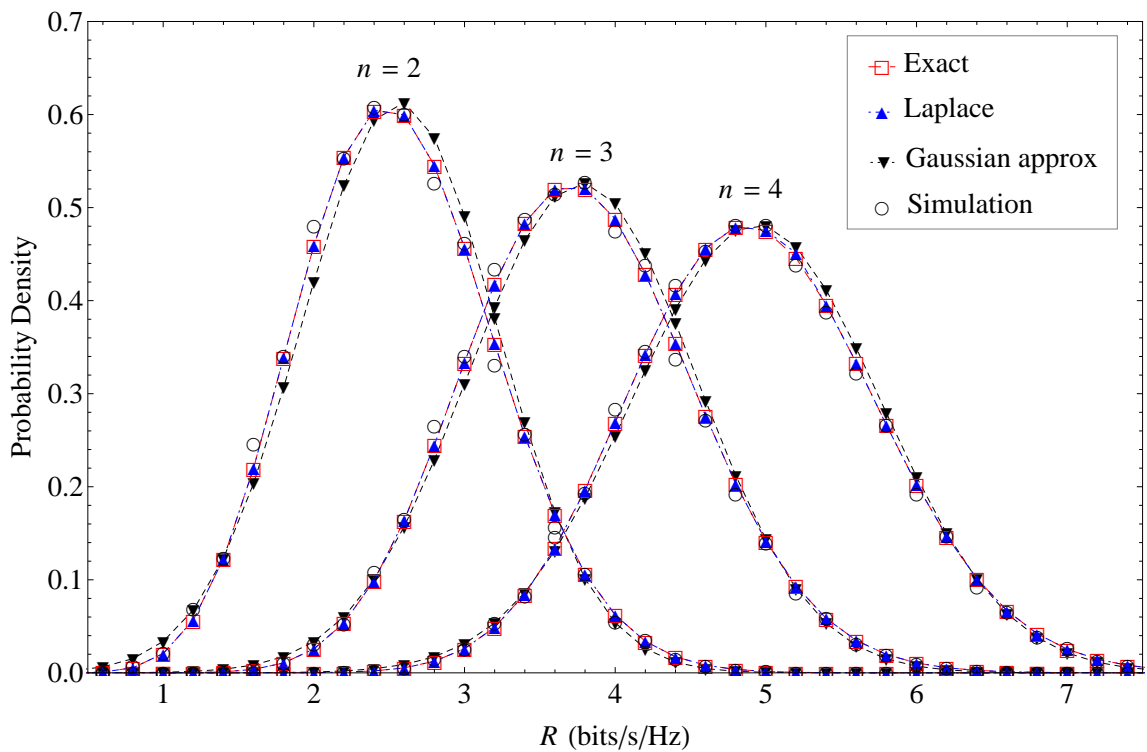}
\caption{Mutual information probability density for $n=\{2,3,4\}$, $n_A=4$, $n_B=5$, $a=1$ and $b=1/3$.}
\label{Probability_Density_new}
\end{figure}

Fig.~\ref{Probability_Density_new} shows the PDF of mutual information for $n=\{2,3,4\}$. Notice that as the number of receiving antennas increases the $\mu_{\mathcal{I}_A}$ also increases. When we double the number of antennas from $n=2$ to $n=4$, the $\mu_{\mathcal{I}_A}$ goes from 2.56 to 4.93, almost a twofold increase. This result is in accordance with the well known result that the slope of the curve increases with $\min(n,n_A)$ \cite{tse2005fundamentals}. Note also that the distributions possess Gaussian-like shapes. For comparison, we have plotted the Gaussian approximation using $\mu_{\mathcal{I}_A}$ and $\sigma_{\mathcal{I}_A}^2$.

\begin{figure}[h!]
\centering
\includegraphics[scale=0.75]{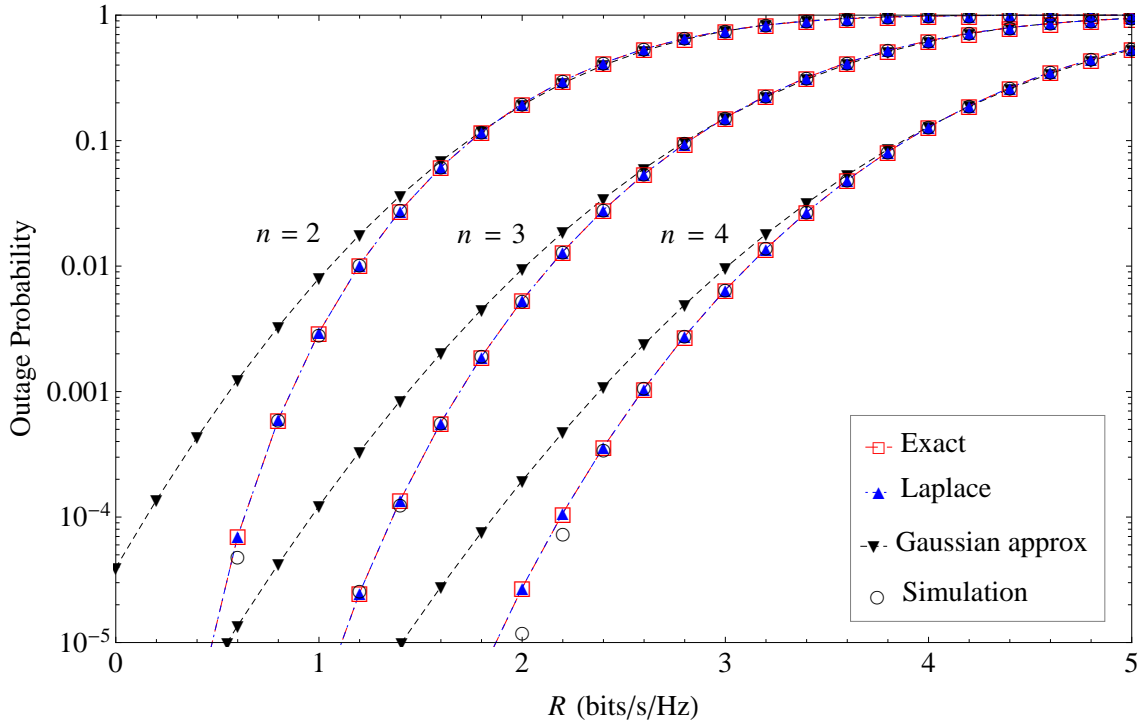}
\caption{Mutual information outage probability for $n=\{2,3,4\}$, $n_A=4$, $n_B=5$, $a=1$ and $b=1/3$.}
\label{Outage_Probability}
\end{figure}

The outage probability is shown in Fig.~\ref{Outage_Probability}. Again, by increasing the number of receiving antennas, the outage probability decreases. For example, for a rate of 3 bits/s/Hz, the outage probability is $\approx 90$\% with $n=2$ antennas, and goes down to less than 1\% for $n=4$ antennas. An outage probability of 1\% allows a bit rate of 1.2 and 2.1 bits/s/Hz with $n=2$ and $n=3$ antennas, respectively. Notice that the Gaussian approximation is indistinguishable for outage probabilities above 10\% for any $n$. For outage probability of 1\%, the error by using this approximation is less than 0.2 bits/s/Hz.

\begin{figure}[h!]
\centering
\includegraphics[scale=0.75]{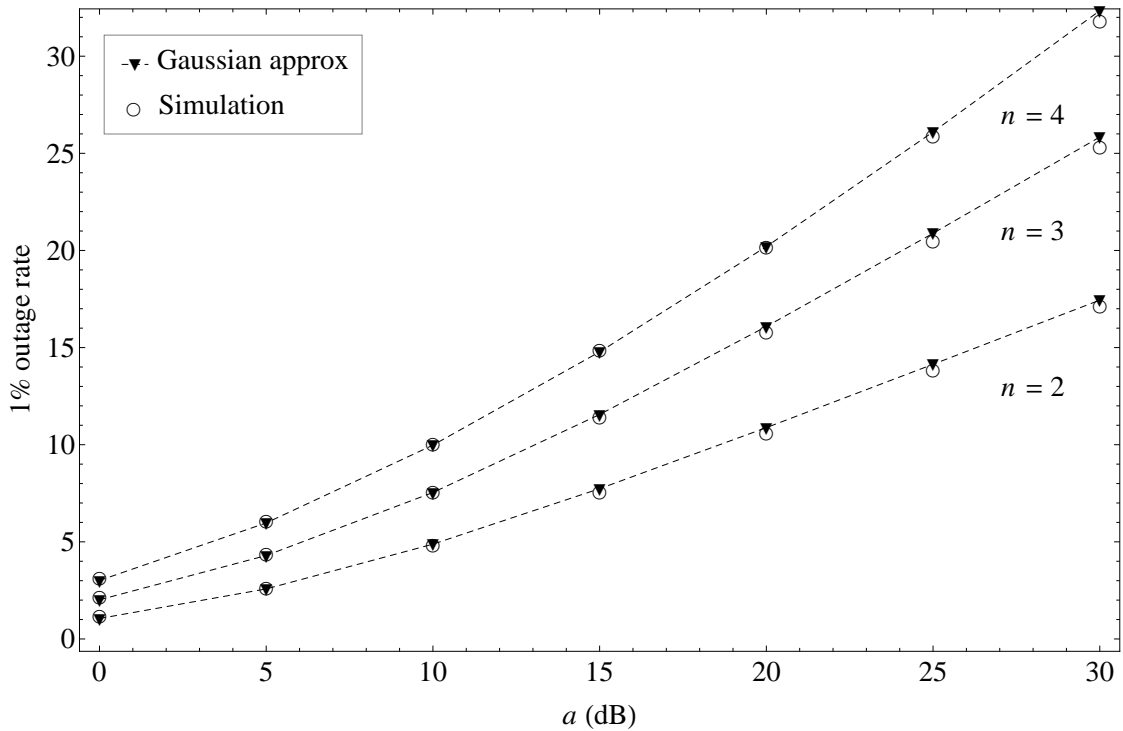}
\caption{1\% outage rate for different $a$ and $n=\{2,3,4\}$, $n_A=4$, $n_B=5$, $b=1/3$.}
\label{1percoutrate}
\end{figure}

We show the outage rate as function of $a$ and $n$ in Fig.~\ref{1percoutrate} for 1\% of outage probability with~$0\leq a \leq 30$~dB, $n=\{2,3,4\}$, $n_A=4$, $n_B=5$, $b=1$. Note that the increase in outage rate is close to linear with $a$. We used only Gaussian approximation to show that, for this purpose, this simpler method presents good results.

\section{Conclusion}

We considered the quotient ensemble involving two Wishart matrices. We worked out exact closed-form expressions for the probability distributions and densities of the extreme eigenvalues. Afterwards, we derived exact expressions for the probability density and outage probability of mutual information of a two-user MIMO MAC network over Rayleigh fading. These expressions allow the analytical evaluation of the probability density and outage beside the current numerical evaluation methods such as Monte Carlo. The exact expressions are presented in two different ways, Laplace transform and by direct integration of joint probability density function of eigenvalues coming from the quotient ensemble.

We showed that the density of mutual information exhibits a Gaussian-like shape. Therefore, besides the exact expressions, we derived expressions to evaluate the mean and variance to invoke the Gaussian approximation method. This approximation method offers a trade-off between complexity and accuracy. For outage probability, the Gaussian approximation shows excellent match with the exact results for outages above 10\%. For lower values, the Gaussian approximation error is relatively small so that we consider the method acceptable due its simplicity of implementation.

Finally, as an example of an application of the derived expressions, we evaluated the effect of the number of receiver antennas in the distribution and outage probability of the receiver. We noted a twofold increase in the mean value of mutual information when we double the number of receiving antennas. On the other hand, the outage rate increased about three times in the low signal to noise ratio regime.

\section*{Acknowledgment}
G. P. and G. F. would like to thank the Capes and CNPq Brazil for supporting this work.

\bibliographystyle{IEEEtran}
\bibliography{library}

\end{document}